\documentclass[runningheads,envcountsame,envcountsect,11pt]{llncs}

\usepackage[english]{babel}
\usepackage[latin1]{inputenc}
\usepackage{amsmath,amssymb,latexsym}
\usepackage{enumerate}
\usepackage{ifthen,calc}
\usepackage{url}
\usepackage{graphics}

\newcommand{\nc}{\newcommand}
\nc{\rnc}{\renewcommand}
\nc{\nev}{\newenvironment}

\nc{\bigmid}{\;\big|\;}
\nc{\Bigmid}{\;\Big|\;}
\nc{\bin}{\textup{bin}}

\nc{\bs}[1]{\boldsymbol{#1}}

\newcommand*{\Next}{\ensuremath{\textrm{\upshape Next}}}

\nc{\lcm}{\ensuremath{\textrm{\upshape lcm}}}
\rnc{\mod}{\ensuremath{\textrm{\upshape mod}}}

\newcommand*{\und}{\ensuremath{\wedge}}

\newcommand*{\oder}{\ensuremath{\vee}}

\newcommand*{\nicht}{\ensuremath{\neg}}

\nc{\FO}{\ensuremath{\textup{FO}}}
\nc{\Bit}{\ensuremath{\textit{Bit}}}

\nc{\uend}{\hfill$\dashv$\par}
\nc{\uendeq}{\eqno\dashv}

\nc{\proofend}{\hfill$\Box$\par}
\nc{\proofendeq}{\eqno\Box}
\rnc{\qed}{\proofend}
\nc{\qedeq}{\proofendeq}

\nc{\bigO}{\ensuremath{{O}}}

\newenvironment{myeqnarray*}{\begin{equation*}\begin{array}{rcl}\displaystyle}{\end{array}\end{equation*}}

\nc{\Randbem}[1]{\marginpar{\footnotesize #1}}

\nc{\ov}[1]{\overline{#1}}
\nc{\vek}{\ov}
\rnc{\vec}{\vek}

\nc{\twodots}{.\,.\,}
\nc{\deff}{:=}
\nc{\parno}{\par\noindent}
\nc{\raus}[1]{}

\nc{\llbrack}{\ensuremath{\lbrack\!\lbrack}}
\nc{\rrbrack}{\ensuremath{\rbrack\!\rbrack}}

\nc{\Blank}{\ensuremath{\Box}}
\nc{\?}{\ensuremath{\circledast}}
\nc{\Nil}{\ensuremath{\circleddash}}
\nc{\undef}{\ensuremath{\bot}}

\nc{\fertig}{\hfill$\Box$\vspace{\topsep}\par}
\nc{\NN}{\ensuremath{\mathbb{N}}}
\nc{\Al}{\ensuremath{\mathbb{A }}}
\nc{\set}[1]{\ensuremath{\{ #1 \}}}
\nc{\setc}[2]{\set{ #1 : #2}}
\nc{\struc}[1]{\ensuremath{\langle #1 \rangle}}

\nc{\lv}{\ensuremath{\textit{lv}}}
\nc{\cp}{\ensuremath{\textit{cp}}}
\nc{\rcp}{\ensuremath{\textit{rcp}}}

\nc{\ind}{\textit{ind}}
\nc{\skel}{\textit{skel}}

\nc{\rev}{\textup{rev}}

\nc{\config}{\ensuremath{\textit{config}}}
\nc{\conf}{\ensuremath{\textit{config}}}
\nc{\tapeconf}{\ensuremath{\textit{tape-config}}}
\nc{\Bin}{\ensuremath{\textit{BIN}}}

\nc{\PartConfT}{\ensuremath{\textit{Conf}_T}}

\nc{\pleft}{\ensuremath{p^{\llbrack}}}
\nc{\pright}{\ensuremath{p^{\rrbrack}}}
\nc{\pact}{\ensuremath{p^{\uparrow}}}

\nc{\phleft}{\ensuremath{\hat{p}^{\llbrack}}}
\nc{\phright}{\ensuremath{\hat{p}^{\rrbrack}}}
\nc{\phact}{\ensuremath{\hat{p}^{\uparrow}}}

\nc{\psleft}{\ensuremath{p'{}^{\llbrack}}}
\nc{\psright}{\ensuremath{p'{}^{\rrbrack}}}
\nc{\psact}{\ensuremath{p'{}^{\uparrow}}}

\nc{\pssleft}{\ensuremath{p''{}^{\llbrack}}}
\nc{\pssright}{\ensuremath{p''{}^{\rrbrack}}}
\nc{\pssact}{\ensuremath{p''{}^{\uparrow}}}

\nc{\Bacc}{\ensuremath{B_{\textit{acc}}}}
\nc{\Brej}{\ensuremath{B_{\textit{rej}}}}

\spnewtheorem{myclaim}{Claim}{\itshape}{\itshape}

    {\hfill$\dashv$\end{list}}

\newsavebox{\fminibox}
\newlength{\fminilength}
\newenvironment{fminipage}[1][\textwidth]{
  \setlength{\fminilength}{#1-4mm-2\fboxrule}%
  \begin{lrbox}{\fminibox}\begin{minipage}{\fminilength}}{
    \end{minipage}\end{lrbox}\noindent\fbox{\usebox{\fminibox}}
}

\newlength{\pwidth}

\renewenvironment{problem}[3]{
  \begin{center}
    \ifthenelse{\equal{#1}{}}%
    {\setlength{\pwidth}{\columnwidth}}%
    {\setlength{\pwidth}{#1cm}}
    \nc{\instance}{\item[Instance]}
    \nc{\parameter}{\item[Parameter]}
    \rnc{\problem}{\item[Problem]}
    \begin{fminipage}[\pwidth]\upshape
      \ifthenelse{\equal{#3}{}}{}{{\scshape #3}}
      \begin{list}{}{
          \ifthenelse{\equal{#2}{}}%
          {\settowidth{\labelwidth}{\textit{Parameter:}}}%
          {\settowidth{\labelwidth}{\textit{#2:}}}
          
          \setlength{\leftmargin}{\labelwidth+\labelsep}
        }
      }{
      \end{list}
    \end{fminipage}
  \end{center}
}

\nc{\ST}{\textsc{ST}}
\nc{\NST}{\textsc{NST}}
\nc{\RST}{\textsc{RST}}
\nc{\DREV}{\textsc{Dreversal}}

\nc{\PTIME}{\textsc{Ptime}}
\nc{\NP}{\textsc{NP}}
\nc{\RP}{\textsc{RP}}
\nc{\co}{\textup{co-}}

\nc{\TIME}{\textsc{Time}}
\nc{\NTIME}{\textsc{NTime}}
\nc{\DTIME}{\textsc{DTime}}
\nc{\RTIME}{\textsc{RTime}}

\nc{\SPACE}{\textsc{Space}}
\nc{\NSPACE}{\textsc{NSpace}}
\nc{\DSPACE}{\textsc{DSpace}}
\nc{\RSPACE}{\textsc{RSpace}}

\nc{\Problem}[1]{\ensuremath{\textsc{#1}}}
\nc{\SETEQUALITY}{\Problem{Set-Equality}}
\nc{\MULTISETEQUALITY}{\Problem{Multiset-Equality}}
\nc{\MSEQUALITY}{\Problem{(Multi)Set-Equality}}
\nc{\CHECKSORT}{\Problem{Check-Sort}}
\nc{\CHECK}{\Problem{Check}}
\nc{\SHORT}{\Problem{Short}}
\nc{\SH}{\Problem{Short-}}
\nc{\PADMSEQ}[1]{{\ensuremath{#1}}\Problem{-Padded-Multi\-set-Equality}}

\nc{\polylog}{\textit{polylog}}

\nc{\LOGSPACE}{\textsc{Logspace}}
\nc{\POLYLOG}{\textsc{Polylogspace}}
\nc{\POLYLOGSPACE}{\POLYLOG}

\nc{\classNC}{\textsc{NC}}
\nc{\NC}[1]{\classNC^{#1}}
\nc{\AC}[1]{\textsc{AC}^{#1}}

\nc{\dcc}[1]{\textup{D}^{#1}}
\nc{\rcc}[1]{\textup{R}^{#1}}
\nc{\pj}{\operatorname{pj}}
\nc{\PJ}{\textup{PJ}}

\begin{document}

\title{Reversal Complexity Revisited}
\author{Andr\'{e} Hernich \and Nicole Schweikardt}
\institute{
   Institut f\"ur Informatik, Humboldt-Universit\"at zu Berlin\\ 
   Unter den Linden 6, D-10099 Berlin, Germany\\ 
   Email: \url{{hernich|schweika}@informatik.hu-berlin.de}
}

\maketitle

\begin{abstract}
We study a generalized version of reversal bounded Turing machines
where, apart from several tapes on which the number of head reversals 
is bounded by $r(n)$, there
are  several further tapes on which head reversals remain unrestricted,
but \emph{size} is bounded by $s(n)$ (where $n$ denotes the input length). 
Recently \cite{grokocschwe05,groschwe05a}, such machines were introduced as
a formalization of a computation model that restricts random access
to external memory and internal memory space. Here, each of the tapes
with a restriction on the head reversals corresponds to an external memory device,
and the tapes of restricted size model internal memory. 
We use $\ST(r(n),s(n),O(1))$ to denote the class of all problems that can be
solved by deterministic Turing machines that comply to the above resource bounds. 
Similarly,  $\NST(\cdots)$ and $\RST(\cdots)$, respectively, are used for the corresponding
nondeterministic and randomized classes.

While previous papers focused on lower bounds for particular
problems, including {sorting}, the {set equality} problem, and
several query evaluation problems, the present paper
addresses the relations between the
$\textsc{(R,N)}\ST(\cdots)$-classes and classical complexity classes and investigates
the structural complexity of the $\textsc{(R,N)}\ST(\cdots)$-classes.
Our main results are
(1) a trade-off between internal memory space and external memory head reversals,
(2) correspondences between the $\textsc{(R,N)}\ST(\cdots)$ 
  classes and ``classical'' time-bounded, space-bounded, reversal-bounded, and circuit complexity
  classes, and
(3) hierarchies of $\textsc{(R)}\ST(\cdots)$-classes in terms of increasing numbers of head
  reversals on external memory tapes.
\end{abstract}

% ==============================================================================

\section{Introduction}
\label{section:Introduction}

Modern software and database technology uses clever heuristics to minimize the
number of accesses to external memory and to prefer \emph{streaming} over
\emph{random accesses} to external memory. There has also been a wealth of
research on the design of so-called \emph{external memory algorithms} (cf.,
e.g.\ \cite{vit01,meysansib03}).  
The classes considered in
\emph{computational complexity theory}, however, usually do not take into
account the existence of different storage media.  

In \cite{grokocschwe05,groschwe05a} complexity classes for
such a scenario were introduced. The two most significant resource bounds in this 
setting are imposed on the
number of random accesses to external memory and the size of the internal
memory. Our complexity classes are based on a standard multi-tape Turing machine.  Some of
the tapes of the machine, among them the input tape, represent external
memory devices. They are unrestricted in size, but access to these tapes is restricted
by allowing only a certain number $r(n)$ (where $n$ denotes the input size) of
reversals of the head directions. This may be seen as a way of (a) restricting the
number of sequential scans and (b) restricting
random access to these tapes, because each random access can be simulated by
moving the head to the desired position on a tape, which involves at most two
head reversals. The remaining tapes of the Turing machine represent the
internal memory. Access to these internal memory tapes (i.e., the number of
head reversals) is unlimited, but their size is bounded by a parameter $s(n)$.
We let $\ST(r(n),s(n),t)$ denote the class of all problems that can be
solved on such an $\big(r(n),s(n),t\big)$-bounded deterministic Turing machine, i.e., a Turing machine
with $t$ external memory tapes which, on inputs of size $n$, performs less than
$r(n)$ head reversals on the external memory tapes, and uses at most space $s(n)$ on the
internal memory tapes.  
Similarly, we use $\NST(r(n),s(n),t)$ and $\RST(r(n),s(n),t)$ for the
corresponding nondeterministic, respectively, randomized classes. The acceptance
criterion for the $\RST(\cdots)$ classes is defined in the same way as for 
the class $\RP$ of \emph{randomized polynomial time}, i.e., ``no''-instances of a
decision problem are \emph{always} rejected, whereas ``yes''-instances are
accepted with probability $\geq 1/2$. 

With the ``external memory'' motivation in mind, we are mainly interested in classes
where the number of head reversals and the internal memory size are comparably small, 
i.e., of size $o(n)$.
Let us, however, emphasize that the main objective in introducing the $\textsc{(N,R)}\ST(\cdots)$-classes
is \emph{not} to provide a ``realistic'' computation model 
suitable for designing efficient external memory algorithms, but to provide 
robust complexity classes which reflect the existence of different
storage media and which are \emph{at least} as powerful as any ``realistic'' external memory
computation model. Thus, \emph{lower bounds} in terms of these complexity classes
will immediately imply lower bounds for the ``external memory complexity'' of certain problems.
For a more detailed discussion on the motivations for considering the $\textsc{(N,R)}\ST(\cdots)$-classes
we refer the reader to the articles \cite{groschwe05a,grokocschwe05,GHS_Randomized_2005} 
and the survey \cite{GKS_FCT05}.

Obviously, our $\ST(\cdots)$-classes are related to the
\emph{bounded reversal Turing machines}, which have been studied in classical complexity theory (see,
e.g., \cite{wagwec86,cheyap91}). However, in bounded reversal Turing machines, the
number of head reversals is limited on \emph{all} tapes, whereas in our
model there is no such restriction on the internal memory tapes. Thus, a priori,
the $\ST(\cdots)$-classes are considerably stronger than conventional bounded reversal classes. 

In \cite{groschwe05a,grokocschwe05,GHS_Randomized_2005}, \emph{lower bounds} for particular
problems, including the {sorting} problem, the {set equality} problem, and
several query evaluation problems, have been shown for the deterministic and randomized
$\ST(\cdots)$ classes. The relations between the
$\textsc{(R,N)}\ST(\cdots)$-classes and  classical complexity classes, as well
as the structural complexity of the $\textsc{(R,N)}\ST(\cdots)$-classes remained as future tasks that
are now addressed by the present paper, whose main results are

\begin{enumerate}[1.]
\item
  a trade-off between internal memory space and external memory head reversals, stating
  that internal memory can be compressed from size $s(n)$ to $O(1)$ at the expense of adding an extra 
  factor $s(n)$ to the external memory head reversals (Theorem~\ref{theorem:internal-memory-to-reversals}).
  \vspace{1mm}
\item
  correspondences between the $\textsc{(R,N)}\ST(\cdots)$ 
  classes and ``classical'' time-bounded, space-bounded, reversal-bounded, and circuit complexity
  classes. For example, we  obtain that 
  $\NP =  \NST(\bigO(1),\bigO(\log n),\bigO(1)) =  \NST(\bigO(\log n),\bigO(1),\bigO(1))$ 
  (Corollary~\ref{cor:NP-characterization}) and that
  $\POLYLOG =   \ST((\log n)^{O(1)},(\log n)^{O(1)},\bigO(1))$
  (Theorem~\ref{thm:POLYLOG-characterisation}).\vspace{1mm}
\item
  hierarchies of $\textsc{(R)}\ST(\cdots)$-classes in terms of increasing numbers of head
  reversals on external memory tapes. E.g., for all $k\geq 2$, we obtain
  $\RST(O(\sqrt[k{+}1]{\log n}),O(\log n),O(1))
    \varsubsetneq  
   \RST(O(\sqrt[k]{\log n}),O(\log n),O(1))
  $ (Theorem~\ref{thm:hier-o}). For the special case where only \emph{one}
  external memory tape is available, we obtain for all functions $r(n)\in o(n/(\log n)^2)$
  that adding one single extra head reversal leads to a strictly larger $\ST(\cdots)$
  class (Theorem~\ref{thm:separation-r(n)->r(n)+1}).
  \vspace{1mm}
\end{enumerate}
\noindent
\textbf{Organization.} In Section~\ref{section:Notation-and-Main-Results} we formally introduce
the $\textsc{(R,N)}\ST(\cdots)$ classes and summarize what has been known about these classes.
Afterwards, in Section~\ref{section:Trade-Off}, we 
show a trade-off between internal memory size and external memory head reversals.
Section~\ref{section:EM-vs-Classical} investigates the relations between the 
$\textsc{(R,N)}\ST(\cdots)$-classes and ``classical'' time-bounded, space-bounded,
reversal-bounded, and circuit complexity classes. In Section~\ref{section:Hierarchies}
we prove hierarchies of deterministic and randomized $\ST(\cdots)$ classes in terms of
increasing numbers of head reversals on external memory tapes.
We close in Section~\ref{section:Conclusion} with a few concluding remarks.

% ==============================================================================

\section{Preliminaries}\label{section:Notation-and-Main-Results}

This section fixes some basic notation, gives a formal introduction of the $\ST(\cdots)$ complexity classes
that were proposed in \cite{groschwe05a,grokocschwe05},
and summarizes what is known about the inclusion structure of these classes.

We write $\NN$ to denote the set of natural numbers excluding 0.
All logarithms are to the base 2 unless otherwise stated.

As our basic model of computation, we use standard multi-tape nondeterministic
Turing machines (NTMs, for short).
The Turing machines we consider will have $t+u$ tapes. We call the first $t$
tapes \emph{external memory tapes} (and think of them as representing $t$
external memory devices); the other $u$ tapes are called \emph{internal memory tapes}. The first
external memory
tape is always viewed as the (read/write) input tape.

Without loss of generality we assume that our Turing machines are
{normalized} in such a way that in each step at most one of its heads
moves to the left or to the right.

A finite run of an NTM $T$ is a sequence $\rho=(\gamma_1,\twodots,\gamma_\ell)$ of 
configurations of $T$ such that $\gamma_1$ is an initial configuration, $\gamma_\ell$ is a
final configuration, and for all $i<\ell$, $\gamma_{i+1}$ can be reached from $\gamma_i$ in a single 
computation step.
 
Let $T$ be an NTM and $\rho$ a finite run of $T$. Let $i\ge 1$ be the number of a tape.
We use 
\(
  \text{rev}(\rho,i)
\)
to denote the number of times the $i$-th head changes
its direction in the run $\rho$.
Furthermore, we let 
\(
\textup{space}(\rho,i)
\)
be the number of cells of tape $i$ that are used by $\rho$.

\begin{definition}[$(r,s,t)$-bounded TM, \cite{groschwe05a,grokocschwe05}]\label{def:boundedTM}
\upshape
  Let $r,s:\mathbb N\to\mathbb N$ and $t\in\mathbb N$. 
  An NTM $T$ is \emph{$(r,s,t)$-bounded}, if every run $\rho$
    of $T$ on an input of length $n$ (for arbitrary $n\in\NN$) satisfies the following conditions:
    \,
(1) $\rho$ is finite, \,
(2) $1+\sum_{i=1}^t\textup{rev}(\rho,i)\le r(n)$\footnote{It is
        convenient for technical reasons to add $1$ to the number
        $\sum_{i=1}^t\textup{rev}(\rho,i)$ of changes of the head direction.
        As defined here, $r(n)$ thus bounds the number of sequential
        scans of the external memory tapes rather than the number of changes
        of head directions.}, \,and \,
(3) $\sum_{i=t+1}^{t+u}\textup{space}(\rho,i)\le s(n)$, where $t+u$ is
      the total number of tapes of $T$. 
\end{definition}

\begin{definition}[$\ST(\cdots)$ and $\NST(\cdots)$ classes, \cite{groschwe05a}]\label{def:ST}
\upshape
  Let $r,s:\mathbb N\to\mathbb N$ and $t\in\mathbb N$. 
  A decision problem belongs to the class $\ST(r,s,t)$ (resp., $\NST(r,s,t)$), if it
  can be decided by a deterministic (resp., nondeterministic) $(r,s,t)$-bounded Turing machine. 
\end{definition}
Note that we put no restriction on the running time or the space used on the
first $t$ tapes of an $(r,s,t)$-bounded Turing machine. The following
lemma shows that these parameters cannot get too large.
\begin{lemma}[{\cite{groschwe05a,GKS_FCT05}}]
\label{lem:time}\label{lemma:tape-length-bound}
  Let $r,s:\mathbb N\to\mathbb N$ and $t\in\mathbb N$, and let $T$ be an
  $(r,s,t)$-bounded NTM. Then for every run $\rho=(\gamma_1,\twodots,\gamma_\ell)$ of $T$ on an input of
  size $n$ we have
  \(
  \ell  \leq  n\cdot 2^{O(r(n)\cdot(t+s(n)))}
  \)
  and thus \ 
  $
    \sum_{i=1}^t\textup{space}(\rho,i) \ \le \  n\cdot  2^{O(r(n)\cdot(t+s(n)))}.
  $
\end{lemma}
In \cite{groschwe05a,GKS_FCT05}, the lemma has only been stated and proved for
deterministic Turing machines, but it is obvious that the same proof also applies
to nondeterministic machines
(to see this, note that, by definition, \emph{every} run of an $(r,s,t)$-bounded Turing machine 
is finite).
\par
In analogy to the definition of \emph{randomized} complexity classes such as the class $\text{RP}$ of
randomized polynomial time (cf., e.g., \cite{baldiagab95,Papadimitriou}), we also consider the randomized version $\RST(\cdots)$ of the
$\ST(\cdots)$ and $\NST(\cdots)$ classes.
The following definition of randomized Turing machines formalizes the intuition that
in each step, a coin can be tossed to determine which particular successor configuration is chosen in this step.
For a configuration $\gamma$ of an NTM $T$, we write 
\(
    \Next_T(\gamma)
\) 
to denote the set of all configurations $\gamma'$ that
can be reached from $\gamma$ in a single computation step.
Each such configuration $\gamma'\in \Next_T(\gamma)$ is chosen with uniform probability, i.e.,
$\Pr(\gamma\to_T \gamma') = 1/|\Next_T(\gamma)|$.
For a run $\rho=(\gamma_1,\twodots,\gamma_\ell)$, the probability $\Pr(\rho)$ that $T$ performs run $\rho$ is
the product of the probabilities $\Pr(\gamma_i\to_T\gamma_{i+1})$, for all $i<\ell$.
For an input word $w$, the probability $\Pr(T\text{ accepts }w)$ that $T$ accepts $w$ is
the sum of $\Pr(\rho)$ for all accepting runs $\rho$ on input $w$.

We say that a decision problem $L$ is solved by
a $(\frac{1}{2},0)$-RTM if, and only if, there is an NTM $T$ such that every run of $T$ has finite length, and 
the following is true for all input instances $w$:
If $w\in L$, then $\Pr(T\text{ accepts } w)\geq \frac{1}{2}$;
if $w\not\in L$, then $\Pr(T\text{ accepts }w) = 0$.

\begin{definition}[$\RST(\cdots)$ classes, \cite{GHS_Randomized_2005}]\label{def:RST}
\upshape
  Let $r,s:\NN\to\NN$ and $t\in\NN$. 
    A decision problem $L$ belongs to the class $\RST(r,s,t)$ if it can be solved by a 
    $(\frac{1}{2},0)$-RTM that is $(r,s,t)$-bounded.
\end{definition}
As a straightforward observation one obtains:
\begin{proposition}\label{prop:DetVsRandVsNdet}
For all $r,s:\NN\to\NN$ and $t\in\NN$, 
$\ST(r,s,t)  \subseteq  \RST(r,s,t) \subseteq \NST(r,s,t)$.
\end{proposition}
For classes $R$ and $S$ of functions we let
\[
    \textup{ST}(R,S,t)  \ :=  \displaystyle\bigcup_{r\in R,s\in S}\!\!\!\textup{ST}(r,s,t)
    \quad \text{and} \quad
    \textup{ST}(R,S,O(1))  \ := \  \displaystyle\bigcup_{t\in\NN}\ \textup{ST}(R,S,t).
\]
The classes $\NST(R,S,t)$, $\RST(R,S,t)$, $\NST(R,S,O(1))$, and
$\RST(R,S,O(1))$ are defined in the analogous way.

As usual, for every complexity class $C$, we write $\text{co-}C$ to denote the class of all 
decision problems whose \emph{complements} belong to $C$. 
Note that the $\RST(\cdots)$-classes consist of decision problems that can be solved
by randomized algorithms that 
\emph{always} reject ``no''-instances and that 
accept ``yes''-instances with probability $\geq \frac{1}{2}$.
In contrast to this, the 
$\text{co-}\RST(\cdots)$-classes consist of problems that can be solved by randomized
algorithms that 
\emph{always} accept ``yes''-instances and that reject ``no''-instances
with probability $\geq \frac{1}{2}$.
\medskip\\
From Lemma~\ref{lem:time}, one immediately obtains
\begin{corollary}\label{cor:time}
  Let $r,s:\mathbb N\to\mathbb N$ and $t\in\mathbb N$. Then
  \[
  \textsc{(N,R)ST}(r,s,t)\ \ \subseteq\ \ \textsc{(N,R)Time}(2^{O(r(n)\cdot
    (t+s(n))) + \log n}),
  \]
  where $\RTIME(2^{O(r(n)\cdot (t+s(n)))+\log n})$ denotes the class of all decision problems that can
  be solved by a $(\frac{1}{2},0)$-RTM that has time bound $2^{O(r(n)\cdot (t+s(n)))+\log n}$.
\end{corollary}
  
\noindent
In particular, whenever \ $r(n)\cdot s(n)\in O(\log n)$, we have
\[
      \ST(r,s,O(1))  \ \subseteq \ \PTIME, \quad
      \RST(r,s,O(1)) \ \subseteq \ \RP, \quad
      \NST(r,s,O(1)) \ \subseteq \ \NP,
\]
where $\RP$ denotes {randomized polynomial time}, i.e., the class of all
 decision problems that can be solved by a polynomial time bounded $(\frac{1}{2},0)$-RTM.

Separation results are known for, e.g.,\ the following decision problems:

\begin{problem}{}{}{\MULTISETEQUALITY}
  \instance $v_1\#\cdots v_m\# v'_1\#\cdots v'_m\#$, \\  where $m\ge 1$ and
  $v_1,\ldots,v_m,v_1',\ldots,v_m'\in\{0,1\}^*$.
\problem Decide if the multisets $\{v_1,\ldots,v_m\}$ and $\{v'_1,\ldots,v'_m\}$ are equal (i.e., they contain 
  the same elements with the same multiplicities).
\end{problem}

\begin{problem}{}{}{\SH\SETEQUALITY}
  \instance $v_1\#\cdots v_m\# v'_1\#\cdots v'_m\#$, \\ where $m\ge 1$ and
  $v_1,\ldots,v_m,v_1',\ldots,v_m'\in\{0,1\}^{2\cdot\log m}$.
\problem Decide if $\{v_1,\ldots,v_m\} = \{v'_1,\ldots,v'_m\}$ as sets (i.e., disregarding
  multiplicities of elements).
\end{problem}

\noindent%
One of the main results of \cite{GHS_Randomized_2005} shows:

\begin{theorem}[\cite{GHS_Randomized_2005}]\label{thm:MultisetEq} 
$\MULTISETEQUALITY$ and $\SH\SETEQUALITY$
\begin{enumerate}[(a)]
\item do not belong to $\RST(r,s,O(1))$, \\
   whenever $r$ and $s$ are functions with
   $r(n) \in o(\log n)$ and 
   $s(n) \in o(\sqrt[4]n/r(n))$. \vspace{1mm}
\item belong to $\NST(3,O(\log n),2)$ and to $\ST(O(\log n),O(1),O(1))$.
\end{enumerate}
Furthermore, the $\MULTISETEQUALITY$ problem 
belongs to $\co\RST(2,O(\log n),1)$. 
\end{theorem}

This immediately leads to the following separations between the deterministic, randomized,
and nondeterministic $\ST(\cdots)$ classes:

\begin{corollary}[\cite{GHS_Randomized_2005}]\label{cor:DetVsRandVsNdet}\mbox{}\\
  Let $r$ and $s$ be functions with 
  $r(n) \in o(\log n)$  and  $s(n) \in o(\sqrt[4]n/r(n))\cap \Omega(\log n)$.
  Then,
  \begin{enumerate}[(a)]
    \item 
        $\RST(O(r),O(s),O(1))\;  \neq \; \co\RST(O(r),O(s),O(1))$, \vspace{1mm}
    \item 
        $\ST(O(r),O(s),O(1)) \; \varsubsetneq \; \RST(O(r),O(s),O(1)) \; \varsubsetneq \NST(O(r),O(s),O(1))$.
  \end{enumerate}
\end{corollary}

% ==============================================================================

\section{Trade-Off between Internal Space and External Head Reversals}
\label{section:Trade-Off}

In this section we show that internal memory tapes can be simulated by
head reversals on external memory tapes:

\begin{theorem}
\label{theorem:internal-memory-to-reversals}
For all $t \in \NN$ and all functions $r$ and $s$ with
$r(n){\cdot} s(n) \in \Omega(\log n)$, we have
\[
 \begin{array}{rcr}
   \ST(r,s,t) & \ \subseteq \ & \ST(\bigO\left(r{\cdot} s\right),\bigO(1),t{+}2), 
  \\[1mm]
   \RST(r,s,t) & \ \subseteq \ & \RST(\bigO\left(r{\cdot} s\right),\bigO(1),t{+}2), 
  \\[1mm]
   \NST(r,s,t) & \ \subseteq \ & \NST(\bigO\left(r{\cdot} s\right),\bigO(1),t{+}2).
 \end{array}
\]
\end{theorem}

\begin{proof}
The proof is inspired by the proof of
Chen and Yap \cite[Theorem~8]{cheyap91}, where it is shown that 
computations in $\DSPACE(s)$ can be simulated by deterministic 2-tape Turing machines
with $O(s)$ head reversals, provided that $s(n)\in \Omega(\log n)$ and
$s$ is \emph{reversal-constructible} (i.e., given $n$ in unary, 
a string of length $s(n)$ can be produced by a multi-tape Turing machine
that performs $O(s(n))$ head reversals).
Note, however, that our following proof of Theorem~\ref{theorem:internal-memory-to-reversals}
does not need an assumption on any kind of reversal-constructibility of $s$ or $r$, 
and that we have to deal with computations
that, in addition to the space-bounded internal memory tapes, also involves
the external memory tapes (which are bounded in the number
of head-reversals, but unbounded in space).

For proving Theorem~\ref{theorem:internal-memory-to-reversals} for the deterministic case,
let $M$ be a deterministic $(r,s,t)$-bounded Turing machine, for functions $r$ and $s$ with
$r(n){\cdot} s(n) \in \Omega(\log n)$.
Without loss of generality, $M$ uses tape alphabet $\set{0,1}$ and has exactly 
one internal memory tape.
We describe a simulation of $M$ by an $(\bigO(r{\cdot}s),\bigO(1),t{+}2)$-bounded 
Turing machine.
Let $x$ be an input of length $n$.

An \emph{internal configuration} of $M$ on input $x$ is a tuple $(q,w_1,w_2)$, 
where $q$ is a state of $M$ and $w_1 w_2$ is a string of length at most $s(n)$,
and describes the situation when $M$ is in state $q$ and contains the string 
$w_1 w_2$ on its internal memory tape, where the head reads the first symbol of
$w_2$.

Our simulation proceeds in \emph{meta-steps}, each of which simulates one step of $M$.
At the beginning, the external memory tapes $t+1$ and $t+2$ contain the same list 
of all possible internal configurations $(q,w_1,w_2)$ of $M$ on input $x$, where 
$w_1 w_2$ has length 1.
The head of tape $t+1$ is placed on the leftmost symbol of the initial internal 
configuration of $M$ on input $x$ in this list, and the head of tape $t+2$ 
is placed at the leftmost symbol of the entire list.
This situation can be achieved with a constant number of reversals.

In the $i$th meta-step, assume that external memory tapes $t+1$ and $t+2$ both contain
the following string: 
\[ 
  \left((C_1\$)^{k+2} \# (C_2\$)^{k+2} \# \ldots (C_k\$)^{k+2} \# \% 
  \right)^{2^\lambda},
 \]
for some $\lambda \in \NN$, where $C_1,C_2,\ldots,C_k$ is an enumeration of all 
internal configurations of $M$ on input $x$, such that there exists an $m\in\NN$ such 
that each configuration has the form $(q,w_1,w_2)$ for some state $q$ and some
string $w_1 w_2$ of length $m$.
We call a substring of the form $(C_1\$)^{k+2}\# 
\ldots (C_k\$)^{k+2}\#\%$ a \emph{segment}, and a substring of the form 
$(C_j\$)^{k+2}$ a \emph{$C_j$-subsegment}.

Assume that the head of tape $t+1$ is placed on the leftmost symbol of the second 
copy of an internal configuration $C_{i-1}$ in the $C_{i-1}$-subsegment of the 
$i$th segment, and that the head of tape $t+2$ is placed on the leftmost symbol of 
the $i$th segment.
By reading $C_{i-1}$ on tape $t+1$ and the symbols currently read on the 
first $t$ external memory tapes, determine the action of $M$ when being in 
this configuration.
If the next internal configuration $C_i$ induced by this action has the form 
$(q,w_1,w_2)$ for some string $w_1 w_2$ of length $m+1$, then modify both 
strings on tape $t+1$ and $t+2$ such that $C_1,\ldots,C_k$ is an enumeration 
of all internal configurations $(q,w_1,w_2)$ of $M$ on input $x$ such that 
$w_1 w_2$ has length $m+1$.
This can be done in a constant number of reversals.
Then, using the remaining $k$ copies of $C_{i-1}$ in the current segment of 
tape $t+1$, search for $C_i$ in the segment of tape $t+2$.
If it has been found, use the remaining copies of $C_i$ in the current segment
of tape $t+2$ to search for $C_i$ in the next segment of tape $t+1$.
If there is no such segment, then duplicate the string on both tapes $t+1$ and 
$t+2$ first.
Finally, simulate $M$'s action on the first $t$ tapes.
This finishes the $i$th meta-step.

Lemma \ref{lemma:tape-length-bound} tells us that on input $x$, the Turing machine $M$ performs at
most $\ell \leq 2^{\bigO(r(n)\cdot s(n))}$ steps.
Therefore, there are at most $\ell$ meta-steps in the simulation.
There are three situations in which a constant number of head reversals occur
during a meta-step:
The first one is that a head on one of the first $t$ tapes changes its 
direction, which happens at most $r(n)$ times.
The second one is that $M$ uses $m$ cells of internal memory before, and 
$m+1$ cells after one step, which happens at most $s(n)$ times.
The last situation is that the string on tapes $t+1$ and $t+2$ is too short,
so that it has to be doubled.
This happens at most $\bigO(\log \ell) = \bigO(r(n){\cdot}s(n))$ times, since there 
are at most $\ell$ steps in the simulation and in each step we use only one 
segment of the string on these tapes. 
In total, there are at most $\bigO(r(n){\cdot}s(n))$ head reversals in the whole 
simulation,
and the proof of Theorem~\ref{theorem:internal-memory-to-reversals} for the 
deterministic case is complete.

The proof for the randomized and nondeterministic case is the same, except that
in each meta-step, the successor configuration $\gamma'$ of the current 
configuration $\gamma$ of $M$ is determined in a randomized way, respectively, in a
nondeterministic way.
Then, in the randomized case, the probability that we choose a successor 
configuration $\gamma'$ of $\gamma$ during the simulation is precisely the 
probability that $\gamma$ yields $\gamma'$ in a run of $M$.
Therefore, the probability that $x$ is accepted in the simulation is the same as
the probability that $x$ is accepted by $M$.
\qed
\end{proof}

\begin{remark}\label{rem:notRevToInt}
Inclusion in the other direction fails in general, at least for the deterministic
and the randomized case.
For example, by Theorem~\ref{theorem:internal-memory-to-reversals} we have
\[
  \Problem{(R)}\ST(\bigO(1),\bigO(\log n),\bigO(1)) \ \subseteq \ 
  \Problem{(R)}\ST(\bigO(\log n),\bigO(1),\bigO(1)).
\]
However, 
\(
  \Problem{(R)}\ST(\bigO(\log n),\bigO(1),\bigO(1))  \nsubseteq  
  \Problem{(R)}\ST(\bigO(1),\bigO(\log n),\bigO(1))
\) 
since, according to Theorem~\ref{thm:MultisetEq}, the $\MULTISETEQUALITY$ problem 
belongs to $\ST(\bigO(\log n),\bigO(1),\bigO(1))$,
but \emph{not} to $\RST(\bigO(1),\bigO(\log n),\bigO(1))$. 
\end{remark}

% ==============================================================================

\section{``External Memory'' vs.\ Classical Complexity Classes}
\label{section:EM-vs-Classical}

In this section we clarify the relations between the $\ST(\cdots)$ classes and
``classical'' complexity classes such as time- or space-bounded classes,
reversal-bounded classes, and circuit complexity classes. 
Figure~\ref{fig:classes} at the end of Section~\ref{section:EM-vs-Classical} visualizes
some of the present section's results.

% ------------------------------------------------------------------------------
\subsection{Time-Bounded Complexity Classes}
\label{section:EM-vs-Classical/NTIME}

Recall that a function $f\colon \NN \rightarrow \NN$ is called \emph{space-constructible} 
(cf., e.g., \cite{Papadimitriou}) if 
there is a deterministic Turing machine that, given $n \in \NN$ in unary on its (read-only)
input tape, writes
$f(n)$ in unary on its output tape and uses $\bigO(f(n))$ space on its work tapes.

\begin{theorem}
\label{theorem:NTIME-in-NST}
For all space-constructible functions $t\colon \NN \rightarrow \NN$, we have
\[
  \NTIME(2^{t(n)}) \ \subseteq \ \NST(3,\bigO(t(n)),2).
\]
\end{theorem}
\begin{proof}
Let $M$ be a nondeterministic Turing machine with time bounded by $2^{t(n)}$.
We describe how to simulate $M$ by a $(3,\bigO(t(n)),2)$-bounded 
nondeterministic Turing machine.
Let $x$ be an input of length $n$.
The machine nondeterministically writes (the same) sequence of configurations 
$C_0,C_1,\ldots,C_k$, $k \leq 2^{t(n)}$, on both external memory tapes.
It does so without any head reversal and by using that $t(n)$ is 
space-constructible.
If $C_k$ is rejecting, then the machine rejects, too.
If $C_k$ is accepting, then the machine checks in a backward scan of both 
external memory tapes that for every $i \in \set{0,1,\ldots,k-1}$, $C_{i+1}$ is 
a successor configuration of $C_i$, and that $C_0$ is the start configuration of
$M$ on input $x$.
The first condition can be checked without any head reversal while scanning 
$C_i$ and $C_{i+1}$ on separate tapes, and the latter one can be checked while 
scanning $C_0$ and $x$ on separate tapes.
The machine accepts if, and only if, both conditions are satisfied.
\qed
\end{proof}

Together with Theorem~\ref{theorem:internal-memory-to-reversals} and
Corollary~\ref{cor:time}, we obtain
the following:

\begin{corollary}
\label{corollary:NTIME-characterisation}
For all space-constructible functions $t(n) \in \Omega(\log n)$, we have
\[
  \begin{array}{rcl}
      \NTIME(2^{\bigO(t(n))}) 
    & \ = \ 
    & \NST(3,\bigO(t(n)),2) 
   \\[0.5ex]
    & \ = \ 
    & \NST(\bigO(1),\bigO(t(n)),\bigO(1))  \ = \ \NST(\bigO(t(n)),\bigO(1),\bigO(1)).
  \end{array}
\]
\end{corollary}
\begin{proof}\mbox{}\\
\mbox{}\qquad\quad
$ 
  \begin{array}[t]{rcl}
     \NTIME(2^{O(t(n))})
   & \ \stackrel{\text{Thm \ref{theorem:NTIME-in-NST}}}{\subseteq} \
   & \NST(\bigO(1),\bigO(t(n)),\bigO(1))
  \\[1ex]
   & \stackrel{\text{Thm \ref{theorem:internal-memory-to-reversals}}}{\subseteq}
   & \NST(\bigO(t(n)),\bigO(1),\bigO(1))
  \\[1ex]
   & \stackrel{\text{Cor \ref{cor:time}}}{\subseteq}
   & \NTIME(2^{O(t(n))}).
  \end{array}
$
\\
\mbox{}\qed
\end{proof}

\begin{remark}\label{rem:}
Corollary~\ref{corollary:NTIME-characterisation} tells us that internal memory space
can freely be traded for external memory head reversals, and vice versa, on 
\emph{nondeterministic} machines. Contrast this with Remark~\ref{rem:notRevToInt}
which states that for \emph{deterministic} and \emph{randomized} machines,
external memory head reversals are strictly more powerful than
internal memory space. I.e.:
\[
 \begin{array}{rcll}
   \NST(\bigO(1),\bigO(\log n),\bigO(1)) & \ = \ &  \NST(\bigO(\log n),\bigO(1),\bigO(1)), & \quad \text{but}
  \\[0.5ex]
   \RST(\bigO(1),\bigO(\log n),\bigO(1)) & \ \varsubsetneq \ & \RST(\bigO(\log n),\bigO(1),\bigO(1)) & \quad\text{and}
  \\[0.5ex]
   \ST(\bigO(1),\bigO(\log n),\bigO(1)) & \ \varsubsetneq \ & \ST(\bigO(\log n),\bigO(1),\bigO(1)). & 
 \end{array}
\vspace{1ex}
\]
\end{remark}

\noindent
Note that Corollary~\ref{corollary:NTIME-characterisation} gives us, in particular, a characterization of the class $\NP$:
\begin{corollary}\label{cor:NP-characterization}
\(
   \NP \ = \ 
   \NST(\bigO(1),\bigO(\log n),\bigO(1)) \ = \ 
   \NST(\bigO(\log n),\bigO(1),\bigO(1)).
\)
\end{corollary}

\begin{remark}
The classes $\PTIME$ and $\RP$ cannot be
characterized in the analogous way: 
Obviously, the $\MULTISETEQUALITY$ problem belongs to $\PTIME \subseteq \RP$ but, due to
Theorem~\ref{thm:MultisetEq}, it does \emph{not} belong to 
$\RST(O(1),O(\log n),O(1)) \supseteq \ST(O(1),O(\log n),O(1))$.
Thus, together with Corollary~\ref{cor:time}  we obtain
\[
     \RP \ \varsupsetneq \ \RST(\bigO(1),\bigO(\log n),\bigO(1)) 
     \quad \text{and} \quad
     \PTIME \ \varsupsetneq \ \ST(\bigO(1),\bigO(\log n),\bigO(1)).
\]
It remains a future task to determine whether the inclusions
\[
     \RP \ \supseteq \ \RST(\bigO(\log n),\bigO(1),\bigO(1)) 
     \quad \text{and} \quad
     \PTIME \ \supseteq \ \ST(\bigO(\log n),\bigO(1),\bigO(1)).
\]
are strict. 
However, even an inclusion of the form
\[
 \PTIME \ \subseteq \ \ST((\log n)^{O(1)},\bigO(1),\bigO(1)) \qquad (*)
\]
seems rather unlikely, since $(*)$, together with the next subsection's 
Theorem~\ref{thm:POLYLOG-characterisation}, would imply that $\PTIME\subseteq \POLYLOGSPACE$
(where $\POLYLOG$ denotes the class of all decision problems
that can be solved in space $(\log n)^{O(1)}$).
\\ 
Currently, the best simulation of time-bounded computations by deterministic reversal-bounded computations seems
to be Li\'{s}kiewicz's result \cite{Liskiewicz} that, for any function $T$ with $T(n)\geq n$,
\(
  \DTIME(T) \subseteq  \ST(O(\sqrt{T}),O(1),O(1)).
\)

\end{remark}

% ------------------------------------------------------------------------------
\subsection{Reversal-Bounded and Space-Bounded Complexity Classes}
\label{section:EM-vs-Classical/Reversals}

Different versions of \emph{reversal-bounded} computation models have been
studied in the literature, cf., e.g., 
\cite{wagwec86,cheyap91,Parberry:TC38-5,Parberry:IPL24-6,Pippenger_FOCS79}.
The model that is closest to our $\ST(\cdots)$ classes certainly being
Chen and Yap's \cite{cheyap91} class $\DREV(r)$ which consists of all decision
problems that can be solved by a deterministic multi-tape Turing machine
that, on an input $x$ of length $n$, performs at most $O(r(n))$ head reversals
on all its tapes, provided that $x$ is a ``yes''-instance of the decision problem
--- on ``no''-instances, Chen and Yap's model does not impose any resource bounds. 
However, as pointed out in \cite{cheyap91}, it does not make any difference
to also impose the resource bounds for ``no''-instances, provided
that the function $r$ is ``reasonable'' in the sense that it is \emph{reversal-constructible},
i.e., there is a deterministic 2-tape Turing machine which, given $n\in\NN$ in unary, generates
$r(n)$ in unary on one of its tapes and uses $O(r(n))$ head reversals in total. It is noted in
\cite{cheyap91} that many complexity functions, including $\log n$, $(\log n)^i$, and $n^i$,
are in fact reversal-constructible.

So, essentially, the class $\DREV(r)$ corresponds
to our class $\ST(O(r),O(1),O(1))$, where only \emph{constant} internal memory 
is available. 
Translated into the present paper's terminology, one of the main results of
\cite{cheyap91} is:

\begin{theorem}[Chen and Yap \cite{cheyap91}]\label{thm:ChenYap}
Let $s$ be a reversal-constructible function and let $r$ be an arbitrary function 
with $r(n),s(n)\in\Omega(\log n)$. Then,
\[
  \DSPACE(s) \ \subseteq \ \ST(O(s),O(1),O(1)) 
  \quad\text{and}\quad
  \ST(O(r),O(1),O(1)) \ \subseteq \ \DSPACE(r^2).
\]
\end{theorem}

\noindent
From this, together with Theorem~\ref{theorem:internal-memory-to-reversals},
one immediately obtains the following:
\begin{corollary}\label{cor:ChenYap:ST2DSPACE}
For all functions $r$ and $s$ with $r(n){\cdot}s(n)\in \Omega(\log n)$, we have
\[
  \ST(r,s,O(1)) \ \subseteq \ \DSPACE(r^2{\cdot}s^2).
\]
\end{corollary}

\noindent
By giving a \emph{direct} proof, we can slightly improve this to

\begin{lemma}
\label{lemma:ST-in-DSPACE}
For all functions $r$ and $s$ with $r(n){\cdot}s(n) \in \Omega(\log n)$, we have
\[
  \ST(r,s,\bigO(1)) \ \subseteq \ \DSPACE(r^2 {\cdot} s).
\]
\end{lemma}
\begin{proof}
Let $r$ and $s$ be functions with $r(n){\cdot}s(n) \in \Omega(\log n)$, 
let $M$ be a $(r(n),s(n),t)$-bounded deterministic Turing machine, and 
let $u$ be the number of internal memory tapes of $M$.
Let $x$ be an input of length $n$.
We describe how to determine the final state of $M$ on input $x$ in space
$\bigO(r(n)^2{\cdot}s(n))$.

The proof idea is to
construct a space-bounded machine that simulates $M$ by re-computing the 
content of a tape cell of $M$ every time $M$ encounters this tape cell.
Here, we proceed by induction on the head reversals.
Assume that we have already constructed a procedure that can compute the 
state and content of $M$'s tapes on input $x$ after 
the $k$th head reversal in space $\bigO(f_k(n))$, for an appropriate function $f_k$.
Then we can compute the state and content of $M$'s tapes after the $(k+1)$st 
head reversal in space $\bigO(f_k(n) + r(n){\cdot}s(n))$ by simulating $M$, starting in 
the state after the $k$th head reversal, where this state and all tape contents
are determined as needed by computing the corresponding bits by the procedure
for $k$. Here, the extra space of size $O(r(n){\cdot}s(n))$ is needed to store
the current head positions on all tapes (recall from 
Lemma~\ref{lemma:tape-length-bound} that $M$ can visit at most $2^{\bigO(r(n)\cdot s(n))+\log n}$
cells on each tape).

More formally, define a function $S$ such that for every input string $y$ and 
every $k \in \NN$, $S(y,k)$ is the state of $M$ on input $y$ directly after the $k$th 
head reversal (if $M$ on input $y$ performs less than $k$ head reversals, then 
$S(y,k) = S(y,k-1)$); $S(y,0)$ is the start state of $M$.
We use two other functions, $P_i$ and $T_i$,
for each tape $i \in \set{1,\ldots,t+u}$ of $M$.
For every input string $y$, every $k \in \NN$, and every $j \in \NN$, $P_i(y,k)$
is the position of the head of the $i$th tape of $M$ on input $y$ after the 
$k$th head reversal, and $T_i(y,k,j)$ is the $j$th symbol on tape $i$ of $M$ on
input $y$ after the $k$th head reversal (again, if $M$ on input $y$ performs less
than $k$ head reversals, then $P_i(y,k) = P_i(y,k-1)$ and $T_i(y,k,j) = 
T_i(y,k-1,j)$); for $k = 0$, $P_i$ and $T_i$ describe the situation at the start
of the computation.

To compute $S(x,k)$, we start simulating $M$ in state $S(x,k-1)$ such that the 
head of tape $i$ is positioned on cell $P_i(x,k-1)$.
Whenever $M$ reads cell $j$ on tape $i$, we compute $T_i(x,k-1,j)$ and continue 
the simulation until $M$ encounters a final state or a head reversal occurs.
Then $S(x,k)$ is the last state of $M$ in the simulation.
$P_i$ and $T_i$ can be computed in a similar way: $P_i(x,k)$ is simply the 
position of the $i$th head at the end of this simulation, whereas $T_i(x,k,j)$ 
is the symbol written into cell $j$ of tape $i$ during the simulation, or 
$T_i(x,k-1,j)$ if this cell has not been passed.
For the simulation we need to maintain the current state and the positions of 
the heads of all tapes.
By Lemma~\ref{lemma:tape-length-bound} these positions can be numbers between 1
and $2^{\bigO(r(n)\cdot s(n))}$ (recall that $r(n){\cdot}s(n) \in \Omega(\log n)$).
Hence, $\bigO(r(n){\cdot}s(n))$ space suffices to store the current state and 
head positions on all tapes.
We additionally need space to compute $S(x,k-1)$, $P_i(x,k-1)$ and $T_i(x,k-1,
j)$.
Since $S$, $P_i$ and $T_i$ can be computed in constant space for $k = 0$, 
it follows by induction on $k$ that $\bigO(k{\cdot} r(n){\cdot} s(n))$ space suffices to 
compute $S$, $P_i$ and $T_i$ for $k \geq 1$.

Finally, to determine the final state of $M$ on input $x$, we iteratively 
compute $S(x,1)$, $S(x,2)$, $\ldots$ until $S(x,k)$ is a final state for some $k$.
Since on an input $x$ of length $n$, $M$ performs at most $r(n)$ head reversals, we have $k \leq r(n)$.
Hence, the final state of $M$ on input $x$ can be determined in space
$\bigO(r(n)^2{\cdot} s(n))$.
\qed
\end{proof}

We write $\POLYLOG$ to denote the complexity class that consists of all decision problems
that can be solved in space $(\log n)^{O(1)}$. 
Using Lemma~\ref{lemma:ST-in-DSPACE}, Theorem~\ref{thm:ChenYap}, and Theorem~\ref{theorem:internal-memory-to-reversals}, 
one easily obtains the following:

\begin{theorem}\label{thm:POLYLOG-characterisation} \hspace{2cm} 
\begin{enumerate}[(a)]
\item 
 $\ST(O(1),O(\log n),O(1)) \subseteq  \LOGSPACE \subseteq \ST(O(\log n),O(1),O(1)) \subseteq \DSPACE((\log n)^2)$\vspace{1mm}
\item 
\(
  \POLYLOG \ = \ \ST((\log n)^{O(1)},\bigO(1),\bigO(1)) \ = \ \ST((\log n)^{O(1)},(\log n)^{O(1)},\bigO(1)).
\)
\end{enumerate}
\end{theorem}
\begin{proof}
 \emph{(a):} \ The first inclusion is due to Lemma~\ref{lemma:ST-in-DSPACE}; the second and the third inclusion are due to
 Theorem~\ref{thm:ChenYap}. \\
 \emph{(b):} \ This immediately follows from  Theorem~\ref{thm:ChenYap} and Theorem~\ref{theorem:internal-memory-to-reversals}.
\qed
\end{proof}

Later on, in Remark~\ref{remark:StrictInclusionInLogspace} we will see that the inclusion of $\ST(O(1),O(\log n),O(1))$ in  $\LOGSPACE$ is, in fact, \emph{strict}.

\begin{remark}
Let us note that our approach for proving Lemma~\ref{lemma:ST-in-DSPACE} does not work
for nondeterministic or randomized machines. In fact, a nondeterministic analogue of
Lemma~\ref{lemma:ST-in-DSPACE} of the form
\[
  \NST(r,s,\bigO(1)) \ \subseteq \ \NSPACE(r^{O(1)} {\cdot} s^{O(1)}) \qquad (**)
\]
seems rather unlikely, since this would imply that $\NP \subseteq \POLYLOG$.
To see this, recall from Corollary~\ref{cor:NP-characterization} that
$\NP = \NST(O(1),O(\log n),O(1))$. Thus, $(**)$ would imply that
$\NP$ is included in  $\NSPACE((\log n)^{O(1)})$ which, due to Savitch's theorem
\cite{Savitch} is equal to $\DSPACE((\log n)^{O(1)})=\POLYLOG$.
\end{remark}

% ------------------------------------------------------------------------------
\subsection{Circuit Complexity Classes}
\label{section:EM-vs-Classical/Circuits}

Recall that $\NC{i}$ is the class of all decision problems that can be solved by
uniform families of circuits of size $n^{O(1)}$ and depth $(\log n)^i$, that 
consist of bounded fan-in $\set{\und,\oder,\nicht}$-gates.
$\AC{i}$ denotes the analogous class where gates of unbounded fan-in are allowed.
Thus, for all $i\geq 0$, we have $\NC{i}\subseteq \AC{i}\subseteq \NC{i+1}$.
Furthermore, $\classNC$ denotes the union of the classes $\NC{i}$, for all $i\geq 0$.
The next lemma shows that 
$(\log n)^{i}$ head reversals and constant internal memory suffice to simulate
all $\NC{i}$ circuits, and that this is optimal in the following sense:
just $o(\log n)$ head reversals are too weak for simulating $\AC{0}$ circuits,
even if internal memory may get as large as $O(\sqrt[4]{N}/\log N)$.

\begin{lemma}\label{lemma:ACi}\label{lemma:ACvsST}
\mbox{}\hspace{4cm}
\begin{enumerate}[(a)]
\item
  $\NC{i} \subseteq \ST(O((\log n)^{i}),O(1),O(1))$, for all $i\geq 0$.\vspace{0.5ex}
\item
  $\SH\SETEQUALITY \in \AC{0}$.\vspace{0.5ex}
\item
  $\AC{0} \not\subseteq \ST(o(\log n),O(\sqrt[4]{N}/\log n),O(1))$.  
\end{enumerate}
\end{lemma}
\begin{proof}
\emph{(a): } It is known that $\NC{i}\subseteq \DSPACE((\log n)^i)$ (cf., e.g.,
\cite[Proposition\;4.2]{BalcDiazGabarro2}).
Due to Theorem~\ref{thm:ChenYap} we have $\DSPACE((\log n)^i)\subseteq \ST(O((\log n)^i),O(1),O(1))$,
and thus the claim follows.
\smallskip\\
\emph{(b): } It is known (cf., \cite{Immerman_Expr}) that
uniform $\AC{0}$ can be characterized as the class of all string-languages that are
definable by a sentence of \emph{first-order logic} with built-in \emph{Bit}-predicate,
$\FO[<,\Bit]$, for short (for an introduction to first-order logic and its relation to 
complexity theory see, e.g., the textbook \cite{Immerman-Buch}). 
To prove \emph{(b)}, it therefore suffices to find an $\FO[<,\allowbreak \Bit]$-sentence $\psi$ that is satisfied
exactly by those strings that are ``yes''-instances of the $\SH\SETEQUALITY$ problem.
To construct such a sentence, we use the result
(cf.\ e.g.\ \cite{FKPS}; or \cite{DLM} for a purely logical proof) that $\FO[<,\Bit]$ has the
``polylog counting capability'', i.e., for every fixed $i\in\NN$ there is an $\FO[<,\Bit]$-formula
$\varphi_{\textit{count}}^i(x,Y)$ such that for all $n\in\NN$, $x\in\set{0,\twodots,n}$, and 
$Y\subseteq \set{0,\twodots,n}$, the structure 
\(
  \big(\set{0,\twodots,n},<,\Bit,x,Y\big) 
\)
satisfies the formula $\varphi_{\textit{count}}^i(x,Y)$ 
if, and only if, $x=|Y| \leq (\log n)^i$.

Recall that an input instance for the $\SH\SETEQUALITY$ problem is a string of the form
\[
   v_1\#\cdots \# v_m\# v'_1\#\cdots \# v'_m\#
\]
where $m\geq 1$ and each $v_i$ and $v'_j$ is a $\set{0,1}$-string of length $2{\cdot} \log m$.
Using the counting formula $\varphi_{\textit{count}}^1(x,Y)$, it is straightforward to 
construct an $\FO[<,\Bit]$-sentence $\psi_{\textit{instance}}$ that is satisfied by exactly those 
strings that are valid input instances
of $\SH\SETEQUALITY$. 
In addition, we can construct an $\FO[<,\Bit]$-sentence 
$\psi_{\textit{yes}}$ that is satisfied 
by a valid input instance if, and only if, this is a ``yes''-instance. To this end, the formula $\psi_{\textit{yes}}$
just has to express that for every position $x$ carrying a $\#$-symbol in the first half, there is an 
according position $y$ in the second half (and vice versa), such that the $\set{0,1}$-strings of length
$2{\cdot}\log m$ to the left of $x$ and $y$, respectively, are equal --- and equality of substrings of length
$2{\cdot}\log m$ can easily be checked when using the counting formula $\varphi_{\textit{count}}^1$.

Finally, the desired $\FO[<,\Bit]$-sentence $\psi$ that defines $\SH\SETEQUALITY$ is chosen as
the conjunction of the two sentences
$\psi_{\textit{instance}}$ and  $\psi_{\textit{yes}}$.
This completes the proof of \emph{(b)}.
\smallskip\\
\emph{(c): } Follows immediately from \emph{(b)} and Theorem~\ref{thm:MultisetEq}.
\qed
\end{proof}

\begin{remark}\label{remark:StrictInclusionInLogspace}
In particular, as an application of Lemma~\ref{lemma:ACi} and Theorem~\ref{thm:MultisetEq}
we obtain that \ $\ST(O(1),O(\log n),O(1)) \varsubsetneq  \LOGSPACE$, because
$\SH\SETEQUALITY$ belongs to $\AC{0}\subseteq\LOGSPACE$, but not to $\ST(O(1),O(\log n),O(1))$.
\end{remark}

In \cite{Pippenger_FOCS79}, Pippenger showed that $\classNC$ is precisely the class of 
languages recognized by deterministic multi-tape Turing machines that 
\emph{simultaneously} have time-bound $n^{O(1)}$ and perform at most $(\log n)^{O(1)}$ head
reversals on all its tapes. The inclusion ``$\subseteq$'' follows by a rather direct construction;
the proof of the opposite inclusion ``$\supseteq$'' is
more intricate and is obtained by showing that a
Turing machine which simultaneously has time-bound $T(n)$ and performs only 
$r(n)$ head reversals, can be simulated by a uniform family of circuits of
depth $\bigO(r(n){\cdot} (\log T(n))^4)$ and size 
$\bigO(r(n){\cdot}T(n)^{O(1)})$. This simulation result was improved by
Parberry \cite{Parberry:IPL24-6} to:

\begin{theorem}[Parberry \cite{Parberry:IPL24-6}]
\label{theorem:SPACE-REV-in-CIRC}
A deterministic $k$-tape Turing machine which simultaneously has space-bound $S(n)$ and
performs at most $r(n)$ head reversals on all its tapes,
can be simulated by a uniform family of circuits (with $\set{\und,\oder,\nicht}$-gates of bounded fan-in)
of depth $\bigO(r(n){\cdot}(\log S(n))^2)$ and width $\bigO(S(n)^k)$.
\end{theorem}

Note that the present paper's $(r,s,t)$-bounded Turing machines, which are used to define the
complexity classes $\ST(r,s,t)$, are different from Pippenger's and Parberry's machines, as the 
latter are \emph{simultaneously} bounded in head reversals and either time or space on
\emph{all} tapes, whereas our $(r,s,t)$-bounded machines have some tapes (namely, the $t$ external
memory tapes) which are unrestricted in size but restricted in head reversals, and some further tapes
(namely, the internal memory tapes) which are restricted in size but unrestricted in terms of head reversals.
However, by using Theorem~\ref{theorem:internal-memory-to-reversals} and 
Lemma~\ref{lemma:tape-length-bound}, Parberry's Theorem~\ref{theorem:SPACE-REV-in-CIRC} immediately
implies that $(r,s,O(1))$-bounded deterministic Turing machines can be simulated by circuits of the
following kind:

\begin{corollary}
\label{corollary:ST-and-circuits}
Let $r$ and $s$ be functions such that $r(n){\cdot} s(n) \in \Omega(\log n)$.
Then, every language in $\ST(r,s,\bigO(1))$ can be decided by a uniform
family of circuits (with $\set{\und,\oder,\nicht}$-gates of bounded fan-in) of depth $\bigO(r(n)^3{\cdot} s(n)^3)$ and size 
$2^{\bigO(r(n)\cdot s(n))}$.
\end{corollary}
\begin{proof}
By Theorem \ref{theorem:internal-memory-to-reversals}, $\ST(r,s,\bigO(1))
\subseteq \ST(\bigO(r{\cdot}s),\bigO(1),\bigO(1))$.
\\
Furthermore,  we know from Lemma~\ref{lemma:tape-length-bound} that an
$(\bigO(r{\cdot}s),\bigO(1),\bigO(1))$-bounded Turing machine uses
space  $\leq 2^{\bigO(r(n)\cdot s(n))}$ on \emph{all} its tapes and therefore simultaneously
has space-bound $S(n)\in 2^{\bigO(r(n)\cdot s(n))}$ and performs at most
$\bigO(r(n){\cdot}s(n))$ head reversals on all its tapes. Thus, Theorem~\ref{theorem:SPACE-REV-in-CIRC}
gives us a uniform family of circuits of depth $\bigO(r(n)^3{\cdot} s(n)^3)$ and width
$S(n)^{O(1)} \subseteq 2^{\bigO(r(n)\cdot s(n))}$.
Since the \emph{size} of a circuit is bounded by the product of its \emph{depth} and its \emph{width},
the claim follows and the proof of Corollary~\ref{corollary:ST-and-circuits} is complete.
\mbox{}\qed\vspace{3ex}
\end{proof}

\begin{figure}[h!]
 \begin{center}
   \includegraphics*[29.5mm,165.1mm][180.8mm,239mm]{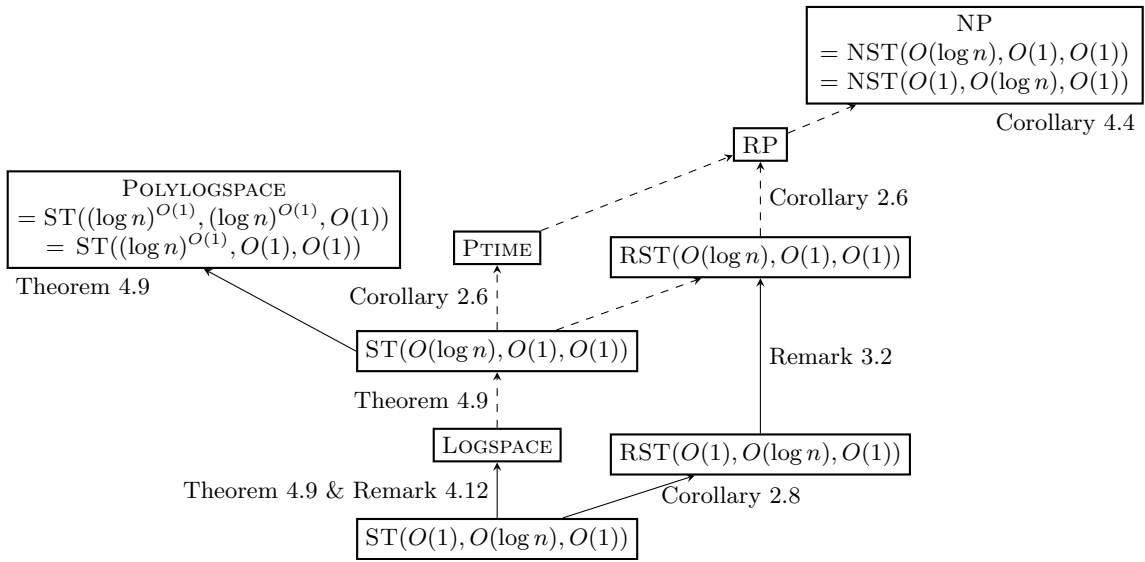}
 \end{center}
 \caption{
   Visualization of some of the relations between $\textsc{(R,N)ST}(\cdots)$ and
   classical complexity classes.
   Solid arrows indicate \emph{strict} inclusions.
 }
 \label{fig:classes}
\end{figure}

% ==============================================================================

\section{Hierarchies of External Memory Classes}
\label{section:Hierarchies}

In this section we prove hierarchies within the $\textsc{(R)}\ST(\cdots)$ classes.
First, in subsection~\ref{section:Hierarchies/Arbitrary-Many-Tapes}, we consider
classes in which arbitrarily many external memory tapes are available; afterwards,
in subsection~\ref{section:Hierarchies/One-Tape}, we  turn to classes with
only one external memory tape.

% ------------------------------------------------------------------------------
\subsection{Hierarchies of Classes with Arbitrarily Many Tapes}
\label{section:Hierarchies/Arbitrary-Many-Tapes}

Chen and Yap have observed in \cite{cheyap91} that their Theorem~\ref{thm:ChenYap}, 
together with the  \emph{space hierarchy theorem} (cf., e.g., the textbook \cite{baldiagab95}), 
leads to a strict hierarchy of reversal complexity classes.
Using Theorem~\ref{theorem:internal-memory-to-reversals}
to shift from internal memory to external head reversals, this leads to:

\begin{proposition}\label{proposition:SpaceHierThm}
Let $r$, $s$, and $R$ be functions such that $r(n){\cdot}s(n)\in\Omega(\log n)$, $R(n)\in\omega(r(n)^2{\cdot}s(n))$, 
and $R$ reversal-constructible and space-constructible. Then, 
\[
   \ST(O(r),O(s),O(1)) \ \varsubsetneq \ \ST(O(R),O(1),O(1)).
\]
\end{proposition}
\begin{proof}\mbox{}\\
\ $\mbox{}\qquad
 \begin{array}[t]{rclll}
    \ST(O(r),O(s),O(1))
  & \ \subseteq \
  & \DSPACE(r^2{\cdot}s)
  & \quad
  & \text{(Lemma~\ref{lemma:ST-in-DSPACE})}
 \\
  & \varsubsetneq
  & \DSPACE(R)
  &
  & \text{(space hierarchy theorem)}
  \\
  & \subseteq
  & \ST(O(R),O(1),O(1))
  &
  &  \text{(Theorem~\ref{thm:ChenYap})}
 \end{array}
$
\\
\mbox{}\qed
\end{proof}

\begin{remark}\label{rem:space-hier}
For example, the above proposition in particular tells us for all $k\geq 1$ that
\[
  \ST(O((\log n)^k),O(\log n),O(1)) \ \varsubsetneq \ \ST(O((\log n)^{2k+2}),O(1),O(1)).
\]
\end{remark}

\noindent
Note that Proposition~\ref{proposition:SpaceHierThm} does not apply to classes where only
$o(\log n)$ head reversals are available. To also treat such cases, we consider 
padded versions of the $\MULTISETEQUALITY$ problem to transfer the lower bound of
of Theorem~\ref{thm:MultisetEq} into the following separations for classes with $o(\log n)$ head reversals:

\begin{lemma}\label{lem:hier-o}
Let $r$ and $s$ be functions such that $r(n)\in o(\log n)\cap \omega(1)$ and
\begin{enumerate}[(i)]
 \item
    given an input string of length $n$, a string of length $r(n)$ can
    be produced by a deterministic $\big(O(r(n)), O(s(n)),O(1)\big)$-bounded Turing machine, and \vspace{0.5ex}
 \item
    given an input of length $\tilde{n}$, 
    a string of length $n$, where $n$ is the smallest number with 
    $r(n)\leq \log \tilde{n} \leq r(n{+}1)$, can be produced by a deterministic $\big(o(\log \tilde{n}),O(\sqrt[5]{\tilde{n}}),O(1)\big)$-bounded Turing machine.
\end{enumerate}
Then, there is a decision problem that belongs to 
$\ST(O(r(n{+}1)),O(s(n{+}1)),O(1))$, but not to  $\RST(o(r(n)),O(\sqrt[5]{2^{r(n)}}),O(1))$.
\end{lemma}
\begin{proof}
Consider the  decision problem

\begin{problem}{}{}{\PADMSEQ{r}}
  \instance $v_1\#\cdots v_m\# v'_1\#\cdots v'_m\#w$, \\  where $m\ge 1$,
  $v_1,\ldots,v_m,v_1',\ldots,v_m'\in\{0,1\}^*$, $w\in \set{1}^*$ such that
  for the 
  total input length $n$ and for $\tilde{n}:= n-|w|$ we have
  $r(n)\leq \log \tilde{n} \leq r(n{+}1)$. 
\problem Decide if the multisets $\{v_1,\ldots,v_m\}$ and $\{v'_1,\ldots,v'_m\}$ are equal.
\end{problem}
To show that $\PADMSEQ{r}$ does 
\[
  \text{\emph{not} belong to }
  \RST(o(r(n)),\allowbreak O(\sqrt[5]{2^{r(n)}}),O(1)),
\]
we give a reduction from  $\MULTISETEQUALITY$  to $\PADMSEQ{r}$:
Given an input of length $\tilde{n}$ of the
form $v_1\#\cdots v_m\#v'_1\#\cdots v'_m\#$ for
the $\MULTISETEQUALITY$ problem, we proceed as follows: First, we
search for a number $n$ such that $r(n)\leq \log \tilde{n}\leq r(n{+}1)$.
Due to assumption \emph{(ii)}, this can be achieved with
$o(\log \tilde{n})$ head reversals on external memory tapes and
$O(\sqrt[5]{\tilde{n}})$ internal memory space.
Afterwards, we pad the input to a string of length $n$ to obtain an input instance for
$\PADMSEQ{r}$. Thus, if $\PADMSEQ{r}$ belonged to 
\(
  \RST(o(r(n)),O(\sqrt[5]{2^{r(n)}}),O(1)),
\)
then 
$\MULTISETEQUALITY$ could be solved by a randomized 
$\big(o(\log \tilde{n}), O(\sqrt[5]{\tilde{n}}),\allowbreak O(1)\big)$-bounded Turing machine 
(to see this, recall that $r(n)\leq \log \tilde{n}$ and
 $\sqrt[5]{2^{r(n)}}\leq \sqrt[5]{2^{\log \tilde{n}}} = \sqrt[5]{\tilde{n}}$).
This, however, is a contradiction to Theorem~\ref{thm:MultisetEq}\,(a).

To see that $\PADMSEQ{r}$ belongs to  $\ST(O(r(n{+}1)),O(s(n{+}1)),\allowbreak O(1))$, we construct a 
Turing machine $M$ which, in a first phase, tests if the input is an admissible instance
for $\PADMSEQ{r}$ as follows:
$M$ checks if the input is of the form $v_1\#\cdots v_m\#v'_1\#\cdots v'_m\#w$, for some $m\geq 1$,
such that $w\in\set{1}^*$ and $v_i,v'_j\in\set{0,1}^*$,
for all $i,j\leq m$ (this can be done with a constant number of head reversals on two
external memory tapes). Then, $M$ checks
for the total input length $n$ and the number
$\tilde{n}:= n-|w|$, that $r(n)\leq \log \tilde{n}\leq r(n{+}1)$ (due to assumption \emph{(i)}, this
can be done with $O(r(n{+}1))$ head reversals on external memory tapes and internal memory
space $O(s(n{+}1))$). This completes $M$'s first phase. Provided that the
input is an admissible instance, $M$ then sorts each of the two  lists  
$v_1\#\cdots v_m\#$ and $v'_1\#\cdots v'_m\#$ of $\set{0,1}$-strings in
ascending lexicographic order. Due to \cite[Lemma\,7]{cheyap91}, this can be achieved
with constant internal memory space and 
$O(\log \tilde{n})\leq O(r(n{+}1))$ head reversals on two external memory tapes. Afterwards,
a single scan of the two sorted lists in parallel suffices to decide if the multisets
$\set{v_1,\twodots,v_m}$ and $\set{v'_1,\twodots,v'_m}$ are equal.
\qed
\end{proof}

\noindent
As a consequence of Lemma~\ref{lem:hier-o} we obtain, for example, the following separation result:

\begin{theorem}\label{thm:hier-o}
For every $k\geq 2$, 
\[
   \ST(O(\sqrt[k]{\log n}),O(\log n),O(1)) \ \not\subseteq \ 
   \RST(o(\sqrt[k]{\log n}),(\log n)^{O(1)},O(1)).
\]
In particular, this implies the following hierarchy of classes for all $k\geq 2$:
\[
   \textsc{(R)}\ST(O(\sqrt[k{+}1]{\log n}),O(\log n),O(1))
   \ \varsubsetneq \ 
   \textsc{(R)}\ST(O(\sqrt[k]{\log n}),O(\log n),O(1)).  
\]  
\end{theorem}
\begin{proof}
The second statement obviously follows from the first statement.
To prove the first statement, use Lemma~\ref{lem:hier-o} for the functions
$r(n)\deff \sqrt[k]{\log n}$ and $s(n)\deff \log n$. Obviously, $r(n)\in o(\log n)\cap \omega(1)$.
Furthermore, assumption \emph{(i)} of Lemma~\ref{lem:hier-o} is satisfied, because
on an input of length $n$, internal memory of size $s(n)=\log n$ is available and thus 
the binary representations of $n$, $\log n$, and $\sqrt[k]{\log n}$ can be computed in
internal memory. Then, a string of length $r(n) = \sqrt[k]{\log n}$ can be produced with
a single scan on an external memory tape. Concerning assumption \emph{(ii)} of Lemma~\ref{lem:hier-o},
we need to find a Turing machine which, on an input of length $\tilde{n}$,
constructs a string of length $n$, where $\sqrt[k]{\log n} = r(n) \leq \log \tilde{n}
\leq \sqrt[k]{\log (n{+}1)}$, i.e., $n\leq 2^{((\log \tilde{n})^k)}\leq n{+}1$.
To achieve this, we may use
internal memory of size up to $\sqrt[5]{\tilde{n}}$.
Note that with  internal memory of that size, the binary representations of numbers of size 
up to $2^{\sqrt[5]{\tilde{n}}} \geq 2^{((\log \tilde{n})^{k})}$ can be handled in internal memory.
Thus, the binary representation of $n$ can be computed in internal memory, and 
afterwards a string of length $n$ can be produced with a single scan on an external memory tape.

Therefore, the assumptions of Lemma~\ref{lem:hier-o} are satisfied, and Lemma~\ref{lem:hier-o}
tells us that 
$   \ST(O(\sqrt[k]{\log n}),O(\log n),O(1)) \ \not\subseteq \ 
    \RST(o(\sqrt[k]{\log n}),\sqrt[5]{2^{r(n)}},O(1)).
$
Theorem~\ref{thm:hier-o} follows, since
$\sqrt[k+1]{\log n} \in o(\sqrt[k]{\log n})$ and $\sqrt[5]{2^{r(n)}} \in 2^{\omega(\log\log n)}$, and
thus is larger than any internal memory bound in $(\log n)^{O(1)}$.
\qed
\end{proof}

\begin{figure}[h!]
  \begin{center}
    \includegraphics*[29.5mm,161.8mm][86.9mm,239mm]{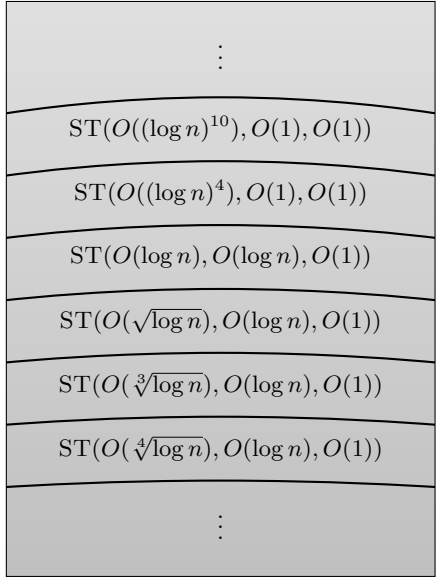}
 \end{center}
 \caption{Visualization of the hierarchy results from Section~\ref{section:Hierarchies/Arbitrary-Many-Tapes}.
 }
 \label{fig:hierarchy}
\end{figure}

\noindent
A visualization of the hierarchy results from
Section~\ref{section:Hierarchies/Arbitrary-Many-Tapes} is shown in Figure
\ref{fig:hierarchy}. 

% ------------------------------------------------------------------------------
\subsection{Hierarchies of Classes with a Single External Memory Tape}
\label{section:Hierarchies/One-Tape}

When considering $\ST(\cdots)$ classes where only \emph{one} external memory tape is available, 
strong results and methods from \emph{communication complexity} can be used
to prove lower bounds.
This technique has already been used in \cite{grokocschwe05} to obtain bounds on the
data complexity of query evaluation problems on data streams and to obtain a 
hierarchy of $\ST(r,s,1)$ classes for the special case where $r$ is
\emph{constant}.
This subsection's main result generalizes the hierarchy of \cite{grokocschwe05} to
non-constant functions $r$. 
The method of choice, again, is to use appropriate
lower bound results for communication complexity.
This subsection's main result is 
\begin{theorem}
\label{thm:separation-r(n)->r(n)+1}
For every logspace-computable function $r$ 
and for all classes $S$ of functions $s\colon \NN \rightarrow \NN$ we have:
\begin{enumerate}[(a)]
\item
  If $r(n)\in o(n/(\log n)^2)$ and \
  $\bigO(\log n) \subseteq S \subseteq o\left(\frac{n}{r(n)\log n}\right)$,
  then \\ $\ST(r(n),S,1) \,\varsubsetneq\, \ST(r(n){+}1,S,1)$. \vspace{1mm}
\item
  If $r(n)^3\in o(n/(\log n)^2)$ and \
  $\bigO(\log n) \subseteq S \subseteq o\left(\frac{n}{r(n)^3\log n}\right)$,
  then \\ $\RST(r(n),S,1) \,\varsubsetneq\, \RST(r(n){+}1,S,1)$.
\end{enumerate}
\end{theorem}

\noindent
Before proving Theorem~\ref{thm:separation-r(n)->r(n)+1}, we introduce some basic
definitions and a strong result from communication complexity. 
Let $X,Y$ be finite sets and $f\colon X \times Y \rightarrow \set{0,1}$ a 
Boolean function.
In Yao's \cite{Yao:STOC79} basic model of communication, Alice gets an input $x
\in X$, Bob gets an input $y \in Y$, and together they wish to evaluate 
$f(x,y)$ by exchanging messages according to a fixed \emph{protocol 
$\mathcal{P}$} that depends on $f$, but not on $x$ or $y$.
The \emph{cost} of $\mathcal{P}$ is the worst case number of bits that have to 
be communicated, so that Alice and Bob both know $f(x,y)$.
\emph{Randomization} can be introduced by giving each player access to his or her own 
random string.
Then, we say that \emph{$\mathcal{P}$ computes $f$ with $1/3$-error} if for all inputs $(x,y) \in
X \times Y$, 
$\mathcal{P}$ 
computes the correct value $f(x,y)$ with probability $\geq 2/3$.

A protocol has \emph{$k$ rounds}, for some $k \in \NN$, if Alice and Bob 
alternately exchange at most $k$ messages.
The \emph{deterministic $k$-round communication complexity of $f$}
is defined as the minimum cost of a deterministic $k$-round 
protocol that computes $f$.
The \emph{randomized $k$-round communication complexity of $f$} 
is the minimum cost of a randomized $k$-round protocol that computes $f$ with 
$1/3$-error.

To prove Theorem \ref{thm:separation-r(n)->r(n)+1}, we consider the 
 \emph{pointer jumping function} $\pj_{\hat{n},k}$ which, for given parameters $\hat{n},k\in\NN$
is defined as follows: \ 
$\pj_{\hat{n},k}(x,y) := 1$ if $x=w_0\cdots w_{\hat{n}-1}$, $y=w_{\hat{n}}\cdots w_{2\hat{n}-1}$, where $w_i \in 
              \set{0,1}^{\log (2\cdot \hat{n})}$, and there exist indices $j_1,\ldots,j_k$ such 
that\footnote{Here, we use $\bin(j)$ to denote the binary representation of $j$ of
length $\log (2{\cdot}\hat{n})$.}
              $w_1 = \bin(j_1)$, $w_{j_i} = \bin(j_{i+1})$ and $w_{j_k}$ has an even number of 
              1s;
and $\pj_{\hat{n},k}(x,y) := 0$ otherwise.
\\
We will use the following lower bounds on the communication complexity of
$\pj_{\hat{n},k}$:

\begin{theorem}[Nisan and Wigderson \cite{NW:JC22-1}]
\label{thm:cc-lower-bounds}
Let $k,\hat{n} \in \NN$.
\\
  If $k < \hat{n}/\log \hat{n}$, then the deterministic $k$-round communication complexity
  of $\pj_{\hat{n},k+1}$ is $\Omega(\hat{n})$. 
 Furthermore,
  if $k < \sqrt[3]{\hat{n}/\log \hat{n}}$, then the randomized $k$-round communication complexity of
  $\pj_{\hat{n},k+1}$ is $\Omega(\hat{n}/k^2)$.
\end{theorem}

\noindent
Theorem \ref{thm:separation-r(n)->r(n)+1} now is obtained as an immediate
consequence of the following lemma, which gives upper and lower bounds for the
language $\PJ_{r+1}$, defined for a given function $r:\NN\to\NN$ via
\[
  \PJ_{r+1} := \setc{\,1^m \# xy \,}{\,
  m\geq 1,\ \text{$x,y\in \set{0,1}^{m\cdot \hat{n}}$, $\hat{n}:=2^{m-1}$, and $\pj_{\hat{n},r(2 m \hat{n}+m+1)+1}(x,y)=1$}\,}.
\]

\begin{lemma}\label{thm:ST1-hierarchy}\hspace{4cm}
\begin{enumerate}[(a)]
\item\label{thm:ST1-hierarchy:item:PJ(r+1)-in-ST1}
  For all logspace-computable functions $r\colon \NN \rightarrow \NN$, we have \\
  \(
    \PJ_{r+1} \ \in \ \ST(r(n){+}1, \bigO((\log r(n))+\log n), 1).
  \) \vspace{1mm}
\item\label{thm:ST1-hierarchy:item:PJ(r+1)-notin-ST1}
  Let $r,s\colon \NN \rightarrow \NN$ such that $r(n) \in o(n/(\log n)^2)$ and 
  $r(n){\cdot}s(n) \in o(n/\log n)$. \\
  Then, 
  \(
     \PJ_{r+1} \ \notin \ \ST(r(n),s(n),1).
  \)\vspace{1mm}
\item\label{thm:ST1-hierarchy:item:PJ(r+1)-notin-RST1}
  Let $r,s\colon \NN \rightarrow \NN$ such that $r(n)^3 \in o(n/(\log n)^2)$ and
  $r(n)^3{\cdot} s(n) \in o(n/\log n)$. \\
  Then, 
  \(
    \PJ_{r+1} \ \notin \ \RST(r(n),s(n),1).
  \)
\end{enumerate}
\end{lemma}
\begin{proof}
\emph{(a):} \
Let $w = 1^m\#w_0 w_1\ldots w_{2^m-1}$ be an input of length $n$, where $w_i\in
\set{0,1}^m$ for all $i \in \set{0,1,\ldots,2^m{-}1}$.
To decide whether $w \in \PJ_{r+1}$, we just have to determine $j_1,j_2,\ldots,
j_k$, $k = r(n)+1$, one after the other, where $\bin(j_i) = w_{j_{i-1}}$ is 
determined from $j_{i-1}$ with at most one head reversal on the external memory
tape.
In internal memory we maintain the value of $m$ ($\log n$ bits), the current value of $j_i$ 
($\log n$ bits), and a counter ($\log r(n)$ bits) which, at the beginning, is 
initialized to $r(n)$ and decreased in every step.
Note that we do not need any head reversal on the external memory tape to 
determine $w_{j_1}$ from $w_0$.
Therefore we need at most $r(n)$ head reversals and $\bigO((\log r(n))+\log n)$
internal memory space.
\smallskip\\
\emph{(b):} \
For contradiction, we assume that there is a deterministic $(r(n),
s(n),1)$-bounded Turing machine $M$ that decides $\PJ_{r+1}$.
Let $Q$ be the set of states of $M$.
For $m \in \NN$ let $\hat{n} := 2^{m-1}$, $n := 2m\hat{n}+m+1$ and $k := r(n)$.
Then, $M$ leads to a deterministic $k$-round communication protocol for 
$\pj_{\hat{n},k{+}1}$ as follows:
Let $x \in \set{0,1}^{m\cdot \hat{n}}$ be the input for Alice and $y \in
\set{0,1}^{m\cdot \hat{n}}$ be
the input for Bob.
Alice simulates $M$ on input $1^m\#xy$ until $M$ reads cell $p := \lvert 1^m\#x
\rvert + 1$ on its external memory tape.
Then, Alice sends the current state of $M$ ($\log \lvert Q\rvert$ bits) and the 
contents of $M$'s internal memory tapes ($O(s(n))$ bits) to Bob.
Bob continues the simulation until $M$ accesses cell $p{-}1$, sends the current 
state of $M$ and the contents of its internal memory tapes to Alice, and so on.
Since $M$ is $(r(n),s(n),1)$-bounded, the head of the external memory tape 
passes cell $p$, or $p{-}1$ respectively, at most $r(n) = k$ times.
So the number of rounds of the protocol is $k$ and the total communication is 
at most $r(n)(O(s(n))+\log \lvert Q\rvert)$, which is of size $o(n/\log n) = o(\hat{n})$ by the 
assumption of \emph{(b)}.
Moreover, $r(n) \in o(n/(\log n)^2) = o(\hat{n}/\log \hat{n})$.
Hence, for sufficiently large $m$ we obtain a $k$-round-protocol for $\pj_{\hat{n},k{+}1}$ with
communication $o(\hat{n})$, where $k < \hat{n}/\log \hat{n}$.
But this contradicts 
Theorem~\ref{thm:cc-lower-bounds}'s statement on deterministic $k$-round communication complexity.
\smallskip\\
\emph{(c):} \ This can be shown 
in the same way as \emph{(b)}, when 
using Theorem~\ref{thm:cc-lower-bounds}'s statement on randomized $k$-round communication complexity.
\qed
\end{proof}

\noindent
Note that Theorem~\ref{thm:separation-r(n)->r(n)+1} is an immediate consequence of Lemma~\ref{thm:ST1-hierarchy}.

% ==============================================================================

\section{Conclusion}
\label{section:Conclusion}

The present paper's main results are
\emph{(a)}~a trade-off between internal memory space and external memory head reversals,
 stating that internal memory can be compressed from size $s(n)$ to $O(1)$ at the expense of adding an extra 
 factor $s(n)$ to the external memory head reversals (see Section~\ref{section:Trade-Off}),
\emph{(b)}~correspondences between the $\textsc{(R,N)}\ST(\cdots)$ 
  classes and ``classical'' time-bounded, space-bounded, reversal-bounded, and circuit complexity
  classes (see Section~\ref{section:EM-vs-Classical}), and
\emph{(c)}~hierarchies of $\textsc{(R)}\ST(\cdots)$-classes in terms of increasing numbers of head
  reversals on external memory tapes (see Section~\ref{section:Hierarchies}).
Visualizations of some of our results can be found in 
Figure~\ref{fig:classes} and Figure~\ref{fig:hierarchy}.

An intriguing future task is to develop techniques for proving lower bounds for appropriate
decision problems in a setting where $\Omega(\log n)$ head reversals and several external memory tapes
are available. Of course, particular separating problems are immediately obtained from the space hierarchy 
theorem via Proposition~\ref{proposition:SpaceHierThm} (see Remark~\ref{rem:space-hier}). However, to find
lower bounds for \emph{``natural''} decision problems can be expected to be rather difficult, since we
know from Theorem~\ref{thm:POLYLOG-characterisation} that $\LOGSPACE \subseteq \ST(O(\log n),O(1),O(1))$.

To conclude the paper, let us comment on the open questions listed in
the conclusion of Chen and Yap's article \cite{cheyap91}. 
One question was whether one can ``speedup'' reversal 
complexity by a constant factor, i.e., whether, for $c>1$, \ $c \cdot r(n)$ reversals can 
be simulated by just $r(n)$ head reversals.
Theorem~\ref{thm:separation-r(n)->r(n)+1} shows that there is no such
speedup for $\ST$ and $\RST$-classes with just \emph{one} external memory tape, as long as 
$r$ is in $o(n/(\log n)^2)$, respectively, in $o(\sqrt[3]{n/(\log n)^2})$.

Another question of \cite{cheyap91} was whether $r(n)$ reversals on two 
external memory tapes can be simulated by
$r(n)^{\bigO(1)}$ reversals on a single external memory tape.
In general, this is not the case.
For example, Lemma \ref{lemma:ST-in-DSPACE} implies that $\LOGSPACE$ is included in
the class $\ST(\bigO(\log n),\bigO(1),\bigO(1))$.
However, it is not included in $\ST(\bigO((\log n)^{\bigO(1)}),\bigO(1),1)$.
To see this,
let $S(x)$ be the subset of $\set{1,\ldots,n}$ represented by the characteristic
string $x$ (i.e., $x = x_1\cdots x_n$ is a bit-string of length $n$, where $x_i =
1$ if and only if $i \in S(x)$).
Let \[L_{Disj} := \set{x\#y \mid \text{$x,y \in \set{0,1}^n$ and $S(x) \cap S(y)
= \emptyset$}}.\]
Then, $L_{Disj}$ is easily seen to be in $\LOGSPACE$. 
On the other hand, \cite[Proposition 4.2]{grokocschwe05} says that $L_{Disj} 
\notin \ST(r(n),s(n),1)$, whenever $r(n){\cdot}s(n) \in o(n)$.
In particular, $L_{Disj} \notin \ST(\bigO((\log n)^{\bigO(1)}),\bigO(1),1)$.

% ==============================================================================

%

\end{document}